\documentclass[fleqn,usenatbib]{mnras}

\usepackage{newtxtext,newtxmath}
\usepackage[T1]{fontenc}
\usepackage{xcolor}

\DeclareRobustCommand{\VAN}[3]{#2}
\let\VANthebibliography\thebibliography
\def\thebibliography{\DeclareRobustCommand{\VAN}[3]{##3}\VANthebibliography}

\usepackage{graphicx}	% Including figure files
\usepackage{amsmath}	% Advanced maths commands
\title[Star formation in magnetized galaxies]{Regulating star formation in a magnetized disk galaxy}

\author[H. Robinson \& J. Wadsley]{
Hector Robinson,$^{1}$
James Wadsley,$^{1}$
\\
$^{1}$Department of Physics and Astronomy, McMaster University, 1280 Main Street West, Hamilton, Ontario, L8S 4M1 Canada
}

\date{Accepted XXX. Received YYY; in original form ZZZ}

\pubyear{2023}

\begin{document}
\label{firstpage}
\pagerange{\pageref{firstpage}--\pageref{lastpage}}
\maketitle

\begin{abstract}
We use high-resolution MHD simulations of isolated disk galaxies to investigate the co-evolution of magnetic fields with a self-regulated, star-forming interstellar medium (ISM). 
The simulations are conducted using the \textsc{Ramses} AMR code on the standard \textsc{Agora} initial condition, with gas cooling, star formation and feedback.  We run galaxies with a variety of initial magnetic field strengths.  The fields evolve and achieve approximate saturation within 500 Myr, but at different levels.  
The galaxies reach a quasi-steady state, with slowly declining star formation due to both gas consumption and increases in the field strength at intermediate ISM densities.  We connect this behaviour to differences in the gas properties and overall structure of the galaxies.  
Stronger magnetic fields  limit supernova bubble sizes.  Different cases support the ISM using varying combinations of magnetic pressure, turbulence and thermal energy. 
 Initially $\gtrsim 1\ \mu G$ magnetic fields evolve modestly and dominate support at all radii.  Conversely, initially weaker fields grow through feedback and turbulence but never dominate the support. This is reflected in the stability of the gas disk.  
This interplay determines the overall distribution of star formation in each case.  We conclude that an initially weak field can grow to produce a realistic model of a local disk galaxy, but starting with typically assumed field strengths ($\gtrsim 1\ \mu G$) will not.

\end{abstract}


%%%%%%%%%%%%%%%%%%%%%%%%%%%%%%%%%%%%%%%%%%%%%%%%%%

%%%%%%%%%%%%%%%%% BODY OF PAPER %%%%%%%%%%%%%%%%%%

\section{Introduction}

%observations of magnetic fields

%Don't need to write intro for my intro - each of these statements should be walked through
%add reference for ubiquitous
Magnetic fields have been detected at all scales in astrophysics \citep{2017Review} and are predicted to play important roles in galaxy evolution. A key question is what role magnetic fields play in regulating galactic star formation. The most straightforward consideration is that magnetic fields provide an additional pressure which can support gas against gravitational collapse on the scale of a galactic disk, but they can also affect galaxy-scale dynamics, the properties of turbulence, the effectiveness of stellar feedback, and the formation of molecular clouds.  Importantly, turbulence acts to amplify and reshape the magnetic field, so in practice we must study the joint evolution of the magnetic field and the ISM together.

%audience should be other grad students - think of people I met at Lyon - primary goal is to maintain interest of people in my field. Do not lose their attention
%don't put pointlessly wordy things in like 'quantum mechanical'
%flip ordering - sycnhrotron first, then paragraph about expected strengths/ordering

% move section about 

%While these methods are not direct measurements of the fields, can be used to infer their presence, and constrain field strengths and morphologies.

Detecting magnetic fields in galaxies is challenging.  A diverse set of methods allows us to infer differing aspects of the field.   These include synchrotron emission, the Zeeman effect, Faraday rotation, polarised thermal emission of magnetically aligned dust grains \citep{pattle}, and polarised emission of starlight due to extinction by aligned dust grains.

Of primary interest is the field strength.  Synchrotron observations can be used to estimate total magnetic field strengths by assuming energy equipartition between magnetic fields and cosmic ray particles. Fields strengths measured in spiral galaxies with this method tend to be around $\sim$10 $\mu$G and decrease slowly with galactocentric radius \citep{fletcher2011,basu,Beck}.  The dependence on the poorly constrained cosmic ray distribution means both the field strength and gradient are highly uncertain.  Due to the diffusive nature and resulting large scale height of cosmic rays, it is expected to probe a thick volume around the galactic disk \citep{zweibel}.  

In the Milky-Way and a few nearby galaxies, line-of-sight magnetic field strengths have also been measured via the Zeeman Effect \citep{crutcher}, which causes emission lines to split when molecules are in the presence of line of-sight magnetic fields. Zeeman observations in the Milky-Way have found fields strengths of $\sim$ 10 $\mu$G at number densities of 10-100 cm$^{-3}$,  In gas at number densities $\gtrsim$ 1000 cm$^{-3}$ the upper envelope to the field strength scales with number density as $B \propto n^{0.5-0.7}$ \citep{crutcher}.  There is considerable spread in these measurements for 10-1000 cm$^{-3}$.  It is commonly assumed that typical field strengths remain flat at lower densities.

The remaining techniques mostly indicate the field morphology. Many observations find large-scale spiral patterns in galactic magnetic fields, even in galaxies that do not have optical spiral structures \citep{chyzy2008,Beck2019,lopez-rodriguez}. 
%magnetic fields tend to be perpendicular with high-column density filaments%
Down to $\sim 100$ pc scales, the fields tend to be aligned parallel with the structure, such as low density filaments \citep{goldsmith2008,sugitani2011}. These results have been confirmed more recently by synchrotron maps of molecular clouds in the Milky way \citep{planck1}. 

%not direct measurement - just relies on fewer assumptions
% not representative of ISM perhaps - gas in supernova remnants 
% put all the simulations papers in the section

%this paragraph may be irrelevant - consider cutting it

There are many theoretical predictions for how magnetic fields should evolve and affect their host galaxies. Tiny cosmic seed fields are amplified exponentially to detectable levels within a few Gyr \citep{lensedgalaxy}. 
On small galactic scales, the turbulent dynamo is expected be ubiquitous.  It can exponentially amplify field strengths over timescales of $\lesssim10$ Myr, saturating at level that is a fraction of the turbulent energy \citep{fed_ssdynamo, Rieder1}.
The $\alpha-\Omega$ dynamo is expected to order the small scale turbulent fields into disk scale regular fields over Gyr timescales \citep{brandenburg2005}. At intermediate scales there is also the gravitational-instability dynamo associated with spiral structures \citep{gidynamo}. Amplification rates in simulations are still highly dependent on numerical resolution and feedback methods \citep{Rieder1}, which can make comparisons between different simulations difficult, however the level they saturate at appears to be independent of those effects.  Thus saturated fields present an appealing target for study that is less dependent on numerical method differences.
%Simulations have found that magnetic field energy densities saturate near a rough equipartition with cosmic rays in the ISM \citep{ponnada_2022}. 

After saturation, the energy density of magnetic fields relative to other sources can vary depending on the phase of gas. The diffuse medium can have significant magnetic support, meaning that the magnetic pressure is comparable to thermal pressure (plasma $\beta$ $~\sim$ 1).
At higher densities, where gas is colder, the primary support is turbulent and the clouds are typically magnetically supercritical (E$_\textrm{mag}$ < E$_\textrm{grav}$ $\sim$ E$_\textrm{turb}$) (see review by \citealt{Beck}). 
% I feel this discussion is quite vague -- it lacks any quantitative values
% On smaller scale turbulence is weaker so magnetic fields might become relatively more important.   However, these scales are probably not resolvable for us?% 
%When combined with the observations of field strengths and morphologies around filaments, this presents a picture where diffuse gas flows along field lines which increases the density but not the field strength, until it becomes supercritical and collapses in star-forming regions \citep{crutcher}.

%that magnetic fields go from being more dynamically important in the diffuse medium, before being overwhelmed by gravity in star-forming regions.
Magnetic fields are expected to play a key role in how gas transitions between phases within the ISM \citep{krumfed_review}. This manifests as a difference in the distribution of gas densities in the ISM, which are created by turbulent compression and expansion. Supersonic turbulence is expected to develop a lognormal probability density function (PDF) that is also seen in observations \citep{kainulainen}. At high densities when the gas becomes gravitationally unstable, it diverges from the lognormal \citep{burkhart2018}. Magnetic pressure narrows the width of the PDF by resisting turbulence's ability to compress gas. Simulations on cloud and kpc scales have shown that this effect can reduce star formation rates by a factor of 2-3 \citep{fed_klessen,padoan2012,girichidis2018,krumfed_review,hix2023,tigressNCR}. Turbulence also plays a role in the magnetic field strength vs. gas density scaling relations. The 0.5-0.7 power law was originally thought to come from gravitational contraction, but cloud-scale simulations by \citet{huabai} retrieve the same scaling without self gravity and the authors argue it comes from turbulent compression instead.  A key takeaway from this discussion is that the role of magnetic fields on smaller scales is complex and under intense study.  Working on slightly larger, galactic scales, allows for a simpler treatment of star formation.

%controlled studies can be more valuable than putting on all the bells and whistles

%Magnetic field strength is also correlated with star formation rate

%dynamos
%no talk of observations after this point - move them earlier
MHD simulations are a powerful tool for studying magnetic fields on many scales. On cosmological scales, magnetic fields are generally not dynamically dominant but some simulations have begun to include them \citep{auriga,fire2,steinwandel3,martin-alvarez2018}. Cosmological seed fields are the origin of all galactic fields; thus cosmological simulations can be used to test models of magnetic field amplification and produce toroidally dominated fields similar to those seen in observations \citep{Rieder3}.  One difficulty with cosmological simulations is that magnetic fields amplify strongly during the poorly resolved and highly chaotic infall and merging phases.   These processes apply ongoing, significant perturbations to the state of the galaxy.  There is also the computational expense of simulating a large cosmological environment for the full history of the universe.

An alternative approach is high resolution simulations of individual galaxies, including the effects of magnetic fields. \citet{Koertgen2019} simulated an isolated disk galaxy and showed that magnetic fields can speed up disk fragmentation and drive outflows hundreds of parsecs above the disk even without stellar feedback. Galaxies that have both star formation and feedback tend to have lower star formation rates when MHD is also included and can magnetize their CGM with magnetic outflows \citep{steinwandel1,steinwandel2,pakmor,wissing}, 
suggesting that MHD can help regulate star formation on disk scales. Recent simulations also suggest that the magnetic field strength continuously declines with density, effectively as a power law, extending to densities below the point where observationally inspired models suggest it should become constant (e.g. \citep{ponnada_2022}).  

Prior work has tended to focus on amplification rates and the final magnetic configuration (e.g. \citealp{Rieder2,su2018}).  However, it is also important consider how magnetic fields affect the state of the gas in the multiphase interstellar medium and whether the star formation is regulated in the same way as we introduce progressively stronger fields.  

In this paper we present a controlled study, simulating an isolated galaxy with a well-known initial condition, with and without magnetic fields, with several different initial field strengths.  Thus we can focus on the development and co-evolution of magnetic fields due to self-regulated star formation and feedback in a galactic disk.   
By running such cases for several dynamical times, we expect to produce a steady-state, self regulated star-forming ISM to study.  In addition, by using a standard setup and simple, well-tested star formation and feedback models, we aim to make the interpretation more straightforward.
% dig at fire -- talk about in conclusions

The remainder of the paper is organized as follows: In section \ref{sec:methods}, we describe the simulation method and magnetized galaxy setup. In section \ref{sec:magfieldevol}, we analyse the evolution of the magnetic field in each galaxy, including the approach to a saturated state and comparing the strengths to observations.  We examine the resulting visual appearance of each galaxy in section~\ref{section:vis}. 
In section \ref{section:sf}, we examine the overall star formation and its radial distribution in each case.  We then explore how this is reflected in their ISM.  Section \ref{gas_properties} examines the gas properties, seeking to determine the underlying drivers of the differences in star formation rates and their connection to the magnetic fields and other forms of support for the gas.  In section \ref{stability}, we study the combined effect of the different support mechanisms on the gravitational stability of the galactic disks.  In section \ref{discussion} we discuss our results and future work. Finally we summarize our conclusions in section \ref{conclusions}. 

%show 
 
%reference links

%SMUGGLE https://arxiv.org/pdf/1905.08806.pdf

%smuggle - arepo successor to effective ism model. 
%cloudy metal cooling, UV background, cosmic ray, PE, self shielding factor,  min T ~ 10K
% epsilon ff = 0.01
% SN energy + momentum split
% photoionisation done probabistically, radiation pressure asdded as momentum, winds

\section{Simulation Method}
\label{sec:methods}
We conduct magnetohydrodynamic (MHD) simulations of isolated galaxies using the adaptive mesh refinement (AMR) code \textsc{Ramses} \citep{RAMSES} to solve the ideal MHD equations using an HLLD approximate Riemann solver \citep{hlld}. The solenoidal constraint ($\nabla \cdot B=0$) is enforced with the constrained transport method \citep{constrained_transport}. 
The dynamics of stars and dark matter are solved using the particle-mesh technique \citep{pm}. Gas cooling and heating is included via the \textsc{Grackle} chemistry and cooling library \citep{grackle}. {\sc Grackle} uses metal cooling rates tabulated from output from the photo-ionization code \textsc{Cloudy} \citep{Cloudy}. We also include a photoelectric heating rate of $\zeta = 4 \times 10^{-26}$ erg cm$^{-3}$ s$^{-1}$, which allows for a two-phase ISM similar to that proposed by \citet{wolfire}.

\subsection{Initial Conditions and Refinement}
\label{sec:ic}
The galaxies we simulate are all based on the medium-resolution isolated disk galaxy from the \textsc{Agora} Project \citep{Agora2016}, but with an initial magnetic
 field added. It has an active dark matter halo ($M_{200} = 1.074\times10^{12}$M$_{\odot}$) that follows an NFW profile, a stellar bulge ($M_{B} = 4.297\times10^{9}$M$_{\odot}$) that follows a Hernquist profile, and a disk ($M_{D} = 4.297\times10^{10}$M$_{\odot}$) that is 80\% stars and 20\% gas by mass with a density profile given by 
 \begin{equation}
        \rho_\textrm{gas}(r,z) = \rho_0 e^{(-r/r_d)} e^{(-|z|/z_d)}
\end{equation}
 With $\rho_0 = M_\textrm{gas}/(4\pi r_d^2 z_d)$, where $r_d=3.432$ kpc and $z_d = 0.1 r_d$. The stellar and dark matter components are modelled with collisionless particles, and the gas is initiated on a \textsc{Ramses} AMR Grid. \cite{Agora2016} contains fulls details about the disk setup.

 This initial condition has been described as similar to 'Milky-Way like' spiral galaxy with a redshift of z$\sim$1.  However, the dark matter halo and rotation curve are quite similar to a $z\sim 0$ large disk galaxy, such as NGC 5055 (the Sunflower galaxy). The initial surface densities of gas and stars are also quite similar to NGC 5055 (as noted by \citealt{benincasaetal2020}).  In this sense it is actually a reasonable proxy for a nearby spiral galaxy.

 The gas disk is initialized inside of a domain 600 kpc on a side that has a base grid of 64$^3$ cells, that is allowed to refine an additional 10 levels which gives a spatial resolution of 9.15 pc at the highest level. A cell will refine if it contains more than 10,000 M$_{\odot}$ of gas or if it contains more than 8 collisionless particles. The disk starts off with $M_\textrm{gas}=$ 8.59$\times$10$^9$ M$_{\odot}$, which will decrease throughout the simulation as gas is converted into stars. The gas is initialized to a temperature of 10,000 K and solar metallicity.  
 
%		Stellar Disk & 3.44e4 & 1e6 \\
%        Bulge & 3.44e4 & 1.25e5 \\
%		Halo & 1.25e6 & 1e6 \\

%at 11 kpc - rho = 4 \times 10^{-25} g/cm^3

%Need to add delayed cooling timescale
Inside the gas disk, we initialize a magnetic field with a morphology that is purely toroidal,  containing no vertical or radial components. The field strength scales with gas density as 
 \begin{equation}
     B = B_0 \left(\frac{\rho}{\rho_0}\right)^{2/3}\label{Binit}
\end{equation} B$_0$ is the value of the magnetic field strength on the midplane at the center of the galaxy.  The magnetic field strength decreases further out in the disk, decreasing by a factor of 10 by 12 kpc. 
Due to the grid initialization process, some initial B values are changed by $\pm$ 10\%. Figure \ref{field_profile} includes radial profiles of the initial magnetic field strength (straight red lines).  

We simulate four galaxies which are identical except for the initial magnetic field, with values for each case summarized in Table \ref{tab:sims}. The first case has zero initial magnetic field, equivalent to being simulated with regular hydrodynamics. The remaining three galaxies are referred to as MHD Weak, MHD Medium, and MHD Strong. In each of them, the field strength initially scales with gas density according to equation~\ref{Binit}, but the constant $B_0$ is modified so that the MHD Medium has fields 10 times stronger than the MHD Weak, and MHD Strong has fields 10 times stronger than MHD Medium.  An increase of 10 times in magnetic field strength corresponds to an increase of 100 times in magnetic energy.

\begin{table}
	\centering
	\begin{tabular}{lrrrr} % four columns, alignment for each
		\hline
		Name & B$_\textrm{ISM}$ & $\beta_\textrm{ISM}$ & B$_0$  \\
		\hline
        Hydro & 0 & $\infty$ & 0 \\ 
        MHD Weak & 0.1 $\mu$G & 800 & 0.85 $\mu$G  \\
        MHD Medium & 1 $\mu$G & 8 & 8.5 $\mu$G  \\
        MHD Strong & 10 $\mu$G & 0.08 & 85 $\mu$G  \\ 
		\hline
	\end{tabular}
     \caption{Summary of initial magnetic properties of each simulation.  B$_\textrm{ISM}$ and $\beta_\textrm{ISM}$ are magnetic field and plasma $\beta$ values, respectively, in the typical ISM (gas density $n \sim 0.25$ cm$^{-3}$).  B$_0$ is the corresponding magnetic field strength in equation~\ref{Binit} (at the geometric center of the galaxy). Otherwise initial magnetic field strengths scale as B$\ \propto\rho^{2/3}$. As a result of this scaling, the plasma $\beta$ (P$_\textrm{thermal}$/P$_\textrm{mag}$) increases with radius.}
     \label{tab:sims}

\end{table}

\subsection{Star Formation and Feedback}

Stars particles are formed stochastically via a Schmidt-law of the form

\begin{equation}
        \frac{d \rho_*}{dt} = \frac{\epsilon_\textrm{ff}\rho}{ \, t_\textrm{ff}} \quad \text{if} \quad \rho > \rho_\textrm{crit}
\end{equation}
where $\rho_*$ is the stellar density, $\epsilon_{\textrm{ff}}$ is the star formation efficiency per free-fall time which we set to $\epsilon_{ff}=0.1$, and $\rho_\textrm{crit}$ is a threshold density which corresponds to a number density of 100 cm$^{-3}$.
%add reference for justification of 0.1

Stellar feedback is entirely supernovae (SN), injected as thermal energy 5 Myr after stars first form with $10^{51}$ erg per 91 M$_{\odot}$ of young stars.  This energy is treated via the delayed cooling model of \cite{delayed_cooling}, which allows unresolved superbubbles to grow correctly by initially treating hot supernova ejecta as unresolved, non-cooling bubbles whose energy is converted to regular thermal energy with a e-folding time of 5 Myr.  At the chosen resolution, other forms of feedback are largely unresolved, as are the dense structures in which they would chiefly operate.

We note that these choices differ from the original \textsc{Agora} simulations \citep{Agora2016}.  In particular, \textsc{Agora} used a low efficiency ($\epsilon_{ff}=0.01$) in dense gas.  This pushes the characteristic time for star formation to $\sim 1$ Gyr, effectively forcing a typical overall galactic star formation rate (SFR), that was the same with or without feedback.  With $\epsilon_{ff}=0.1$, stellar feedback is necessary to regulate star formation to the level expected for typical disk galaxies \citep{agertz2015,semenov2018,msc_thesis}. Feedback is expected to couple in such a way to reproduce large scale ISM properties including the scale height \citep{ostriker2010,benincasa2016}. We choose this simple, yet robust approach to star formation and feedback because it is easy to reproduce and well-tested, allowing us to focus on the effects of MHD.

Each galaxy was evolved for 1 Gyr, corresponding to a handful dynamical times at the outer radius.  The inner regions were expected to evolve rapidly and then settle into a quasi-steady state of ongoing star formation.  Thus the goal was to produce an interval of several 100 Myr to study this relatively quiet, self-regulated state in each case.

\section{Simulation results}

At the start of the simulation, all four galaxies begin their evolution by settling from the slightly unstable initial state by compressing vertically. While this happens cold-phase gas condenses and fragments along spiral arms into individual clouds. In the Hydro and MHD Weak  cases, this collapse is violent and results in a starburst, mostly localized to the galactic center (r < 2 kpc). 
In the cases with stronger fields, the fields resist the compression, preventing a starburst from occurring, and the onset of star formation is delayed. As a result, the Hydro and MHD Weak galaxies have a smaller fraction of gas remaining in the galactic center for the remainder of the simulation, for this reason we exclude the center 2 kpc from some of our analysis to ensure a fair comparison between the galaxies.

\subsection{Magnetic Field Evolution}\label{sec:magfieldevol}

\begin{figure*}
	\includegraphics[width=2.1\columnwidth]{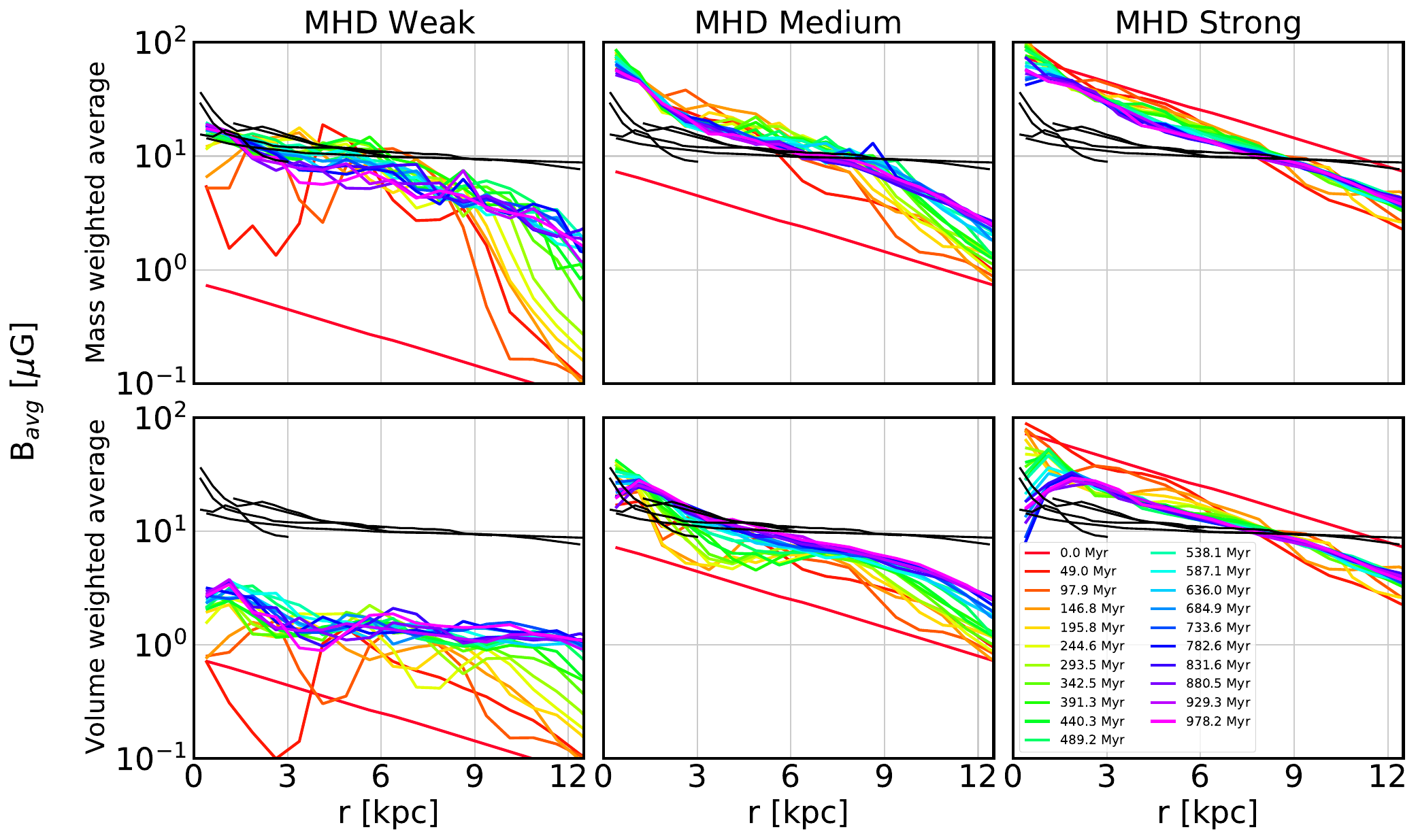}
    \caption{Magnetic field strength vs. galactocentric radius in each MHD galaxy. Top row shows a mass-weighted average, and bottom row shows a volume-weighted average. Color shows the the time of the snapshot from 0 (red), to 1 Gyr (pink). Black lines are data from nearby galaxies from \citet{basu}. Field strengths are calculated in a disk with a height of 1 kpc.}   
    \label{field_profile} 
\end{figure*}

The primary focus of this work is how magnetic fields influence the structure of the ISM and the regulation of star formation at later stages.  Before examining that we give a brief overview of the magnetic field evolution leading to the final state.

{Figure \ref{field_profile} shows the average magnetic field strength vs. radius in a disk of height 1 kpc} for each galaxy over time, and compares them to field strengths from synchrotron observations by \citet{basu}.  Firstly, we note that this observational sample includes NGC 5055, for which the simulation set-up is a reasonable proxy.   Secondly, the variation among the observational sample is also quite small, with estimated field strengths of $\sim$ 10-20 $\mu$G in the radii of interest, 2-10 kpc. 

The figure includes two methods of calculating the average field strength. The first is a mass-weighted average (top panels) and the second is a volume weighted average (lower panels). The mass weighted average gives an estimate of the field strengths in high-density regions, while the volume weighted average probes more typical ISM densities. The method of averaging does not particularly affect the field  estimate in the MHD Strong galaxy because the gas scale height is quite large, and there is very little high density gas created. (see ~\ref{gas_properties} for a quantitative comparison).  For weaker fields, the galaxies are thinner and the differences more pronounced.  

The volume weighted average should be a fair comparison to the synchrotron estimates because synchrotron emission arises from cosmic rays which are typically assumed to have a large scale height \citep{zweibel}.  Many assumptions are required to calculate field strengths from synchrotron emission, such as energy equipartition and assumed cosmic ray energy distributions \citep{basu}. There are also uncertainties in the volume of the synchrotron emitting disk. Most of the assumptions go into the amplitude of the field strength, meaning there is quite a bit of uncertainty in the real field strength. However, the radial profile depends only on the synchrotron intensity and therefore has fewer assumptions built in. The MHD Weak galaxy is promising in this case because the volume weighted average achieves a flat radial profile like the observations. This makes the MHD Weak case the most compelling. In that case, the fields saturate around 1 $\mu$G for most of the disk, whereas the observations estimate 10 $\mu$G. It may be worth revisiting if the observations are overestimating the field strengths because of some of the assumptions built in. This also suggests that initial fields must be well below equipartition in order to naturally evolve to a realistic saturated state that reproduces the radial profile.

%\citet{pakmor} also compare their cosmological galaxies to the same observations using a volume weighted average and get a variety of field strengths whose profiles most closely match our MHD Medium galaxy, having stronger fields than the observations in the center and weaker in the outer disk.

Similar amounts of magnetic field amplification are visible in both the MHD Weak and MHD Medium galaxies regardless of the choice of weighting. The MHD Strong galaxy experiences a net loss of field strength over time, confirming that it was oversaturated from the beginning. Both the MHD Weak and Medium galaxies appear close saturation after about 500 Myr, but they do not saturate at the same value. Field strengths may decrease over time due to magnetic flux leaving the disk. In the inner regions of the MHD Weak galaxy, field strengths peak at around t=500 Myr and then decrease slightly due to flux leaving in vertical outflows. In the MHD Strong galaxy, net flux loss is expected due to magnetic braking.

In all cases, the amplification rate is lower in the outer disk, where there is less star formation (see section \ref{section:sf}).  This behaviour is most consistent with a turbulent style dynamo, both in terms of the rate of growth and the strong association with feedback from star formation.

\begin{figure}
	\includegraphics[width=\columnwidth]{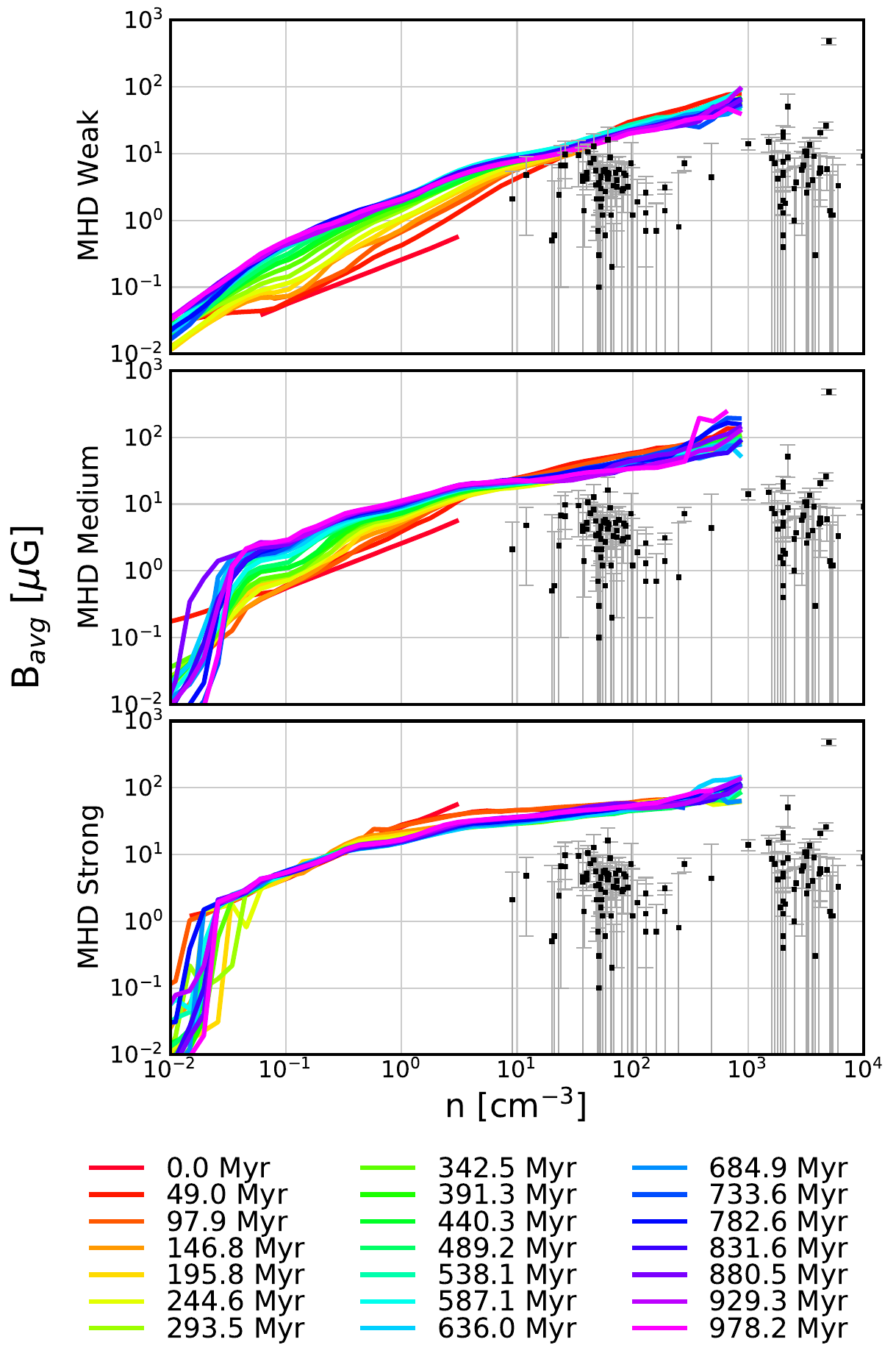}
    \caption{Median magnetic field strength vs. gas number density over time. Black points are measured values from clouds inside the Milky Way \citep{crutcher}. Amplification is visible in diffuse gas in galaxies with weaker initial fields.}
    \label{crutcher_plot}
\end{figure}

Our field strengths can also be compared to those inferred from Zeeman measurements, by placing them on a field strength vs. density (B vs. n) plot, shown in figure \ref{crutcher_plot}.   Zeeman measurements typically probe dense, mostly neutral and molecular gas that is not present in these simulations (see section~\ref{gas_properties}).  In addition, processes such as turbulent ambipolar diffusion are expected to remove field in dense neutral gas.  Thus the numerical results are best interpreted as upper bounds for densities, $n \gtrsim 30$  cm$^{-3}$.  
We plot the median field strength in each density bin, but variations of an order of magnitude are common at a given density. In this figure, field amplification due to dynamo action is seen as a vertical translation. When field strengths increase due to gas compression (or expansion), the points should also move to the right as the density increases (or decreases, respectively). This largely explains the rapid expansion of the plot upward in density and field strength after the initial state.   In the highest density gas field strengths remain similar across all three simulations.  
This is likely due to the fact that the field strength in the dense gas immediately saturates. This dense gas is selectively where star formation occurs, driving turbulence which would feed a rapid, small scale dynamo  (See figure \ref{pressure}). 
Amplification is clearly visible in the diffuse medium with number densities of 0.1-10 cm$^{-3}$. There are differing levels of amplification seen in the diffuse medium in each case.
The 2/3 power-law slope is maintained in the diffuse medium as the field evolves.  There is a shallower power-law slope for intermediate density gas (1-100~cm$^{-3}$), which is flatter in the stronger field cases.  The stronger field forces gas flows  along field lines, limiting field increases due to density changes. 
%The field strengths in the dense gas are mostly independent of the initial state, all three ending up with values of around 100 $\mu$G at a number density of 10$^3$ cm$^{-3}$.

At low densities, the simulated field strengths continue to decrease with density, showing no hint of the constant 10~$\mu G$ field extrapolation below number densities of $\sim 10$ cm$^{-3}$ assumed by, e.g. \citet{crutcher}. 
%discuss magnetic field strength over time, amplification, saturated vs. not saturated

\subsection{Overall structure}
\label{section:vis}

\begin{figure*}
	\includegraphics[width=2\columnwidth]{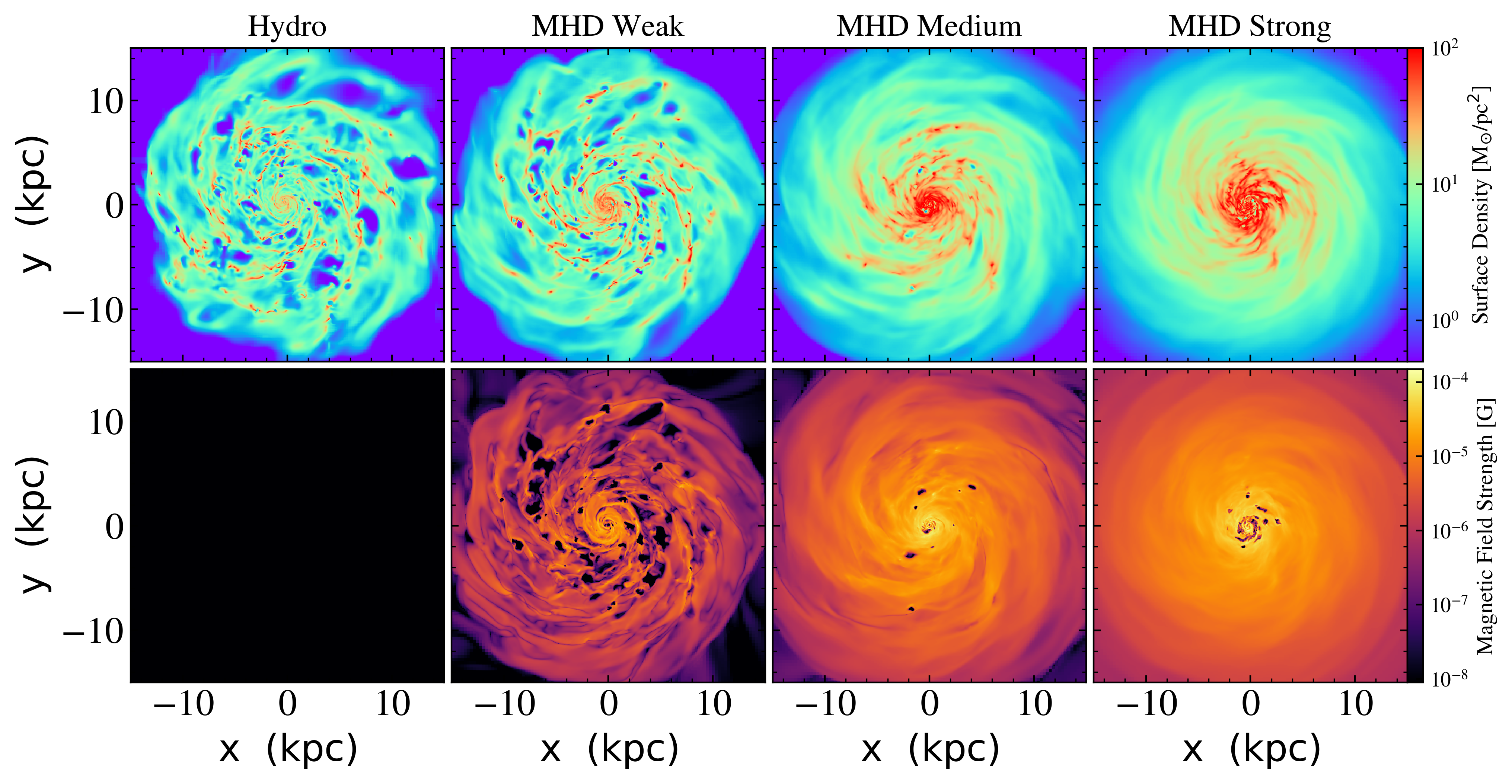}
    \caption{Visualisations of each galaxy 1 Gyr. The top row shows surface density projections, and the bottom row shows slices of magnetic field strength in the midplane. The Hydro case is black because there is zero magnetic fields everywhere.}
    \label{projection}
\end{figure*}

We now proceed with an examination of the state of the galaxies after 1 Gyr of evolution.  Figure \ref{projection} shows face-on visualisations of each galaxy. The top row shows the surface densities of gas in each galaxy. The excess of gas in the medium and strong galaxies is clearly visible. Those galaxies also appear less fragmented, with more flocculent spiral arms and fewer superbubble holes. 
The bottom row shows slices of the magnetic field strength in the midplane of the galactic disk. Although there is some amplification of the fields during the galaxies evolution, the fields do not saturate at the same level and the case with the stronger initial fields still has the strongest magnetic fields at the end of the run. The stronger magnetic fields have less structure in them, mostly due to them having lower star formation rates (see section~\ref{section:sf}), but also due to the weaker fields being less dynamically important and having less resistance to being pushed around by motions of the gas. 
Large scale spiral structure is reflected in the magnetic fields, and the stronger field cases result in spiral morphology that is less disrupted by superbubbles.  To understand this better, we need to examine the distribution of star formation.

\begin{figure}
	\includegraphics[width=\columnwidth]{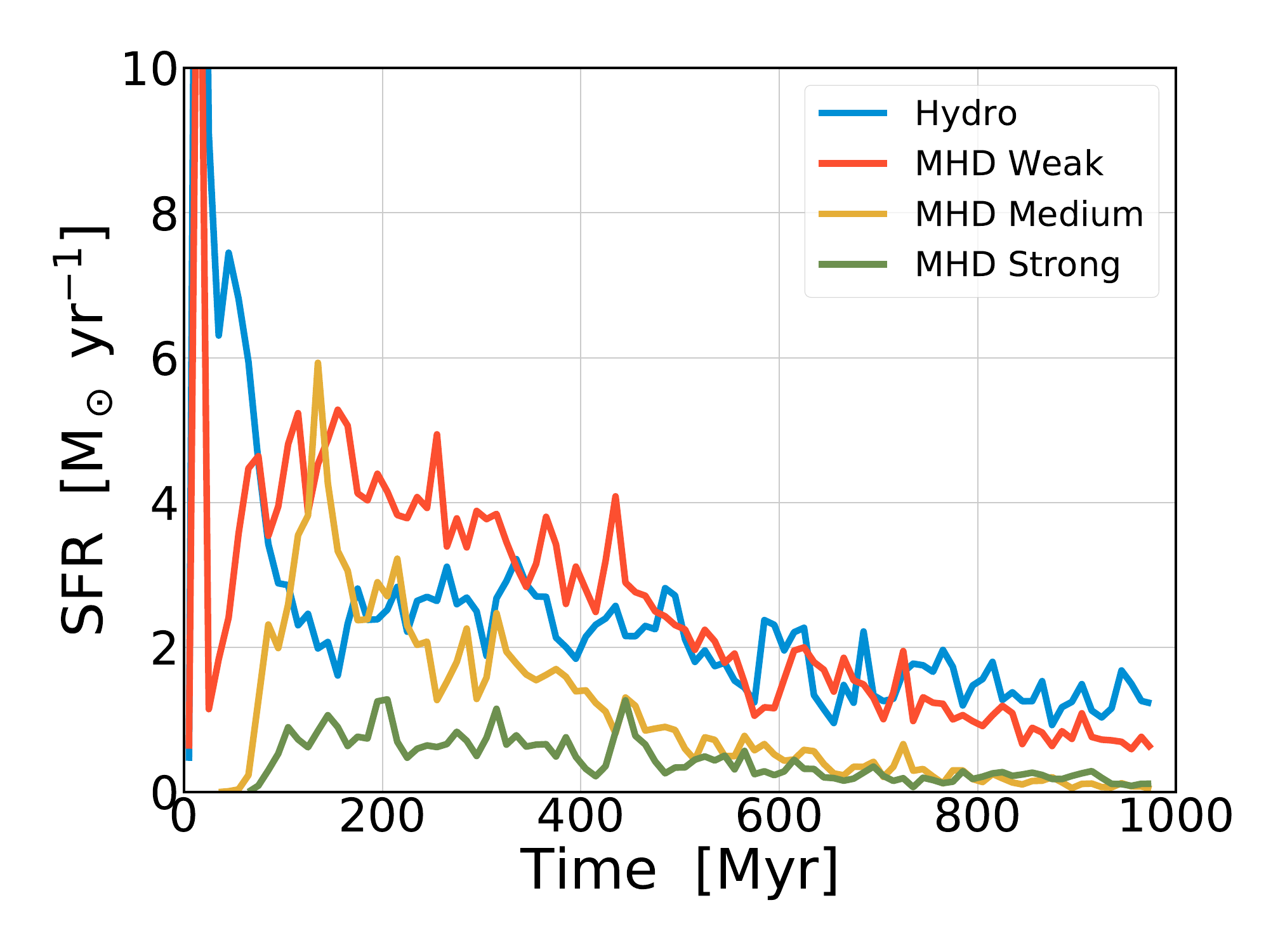}
    \caption{Star formation rate vs. time for each galaxy, excluding the central 2 kpc. Star formation rate is calculated by summing the mass of all stars formed within each time bin and dividing by the size of the bin. For the MHD Weak and Medium galaxies, field strengths decline over time due to increasing field strengths.}
    \label{sfr}
\end{figure}
\begin{figure}
    \centering
	\includegraphics[width=\columnwidth]{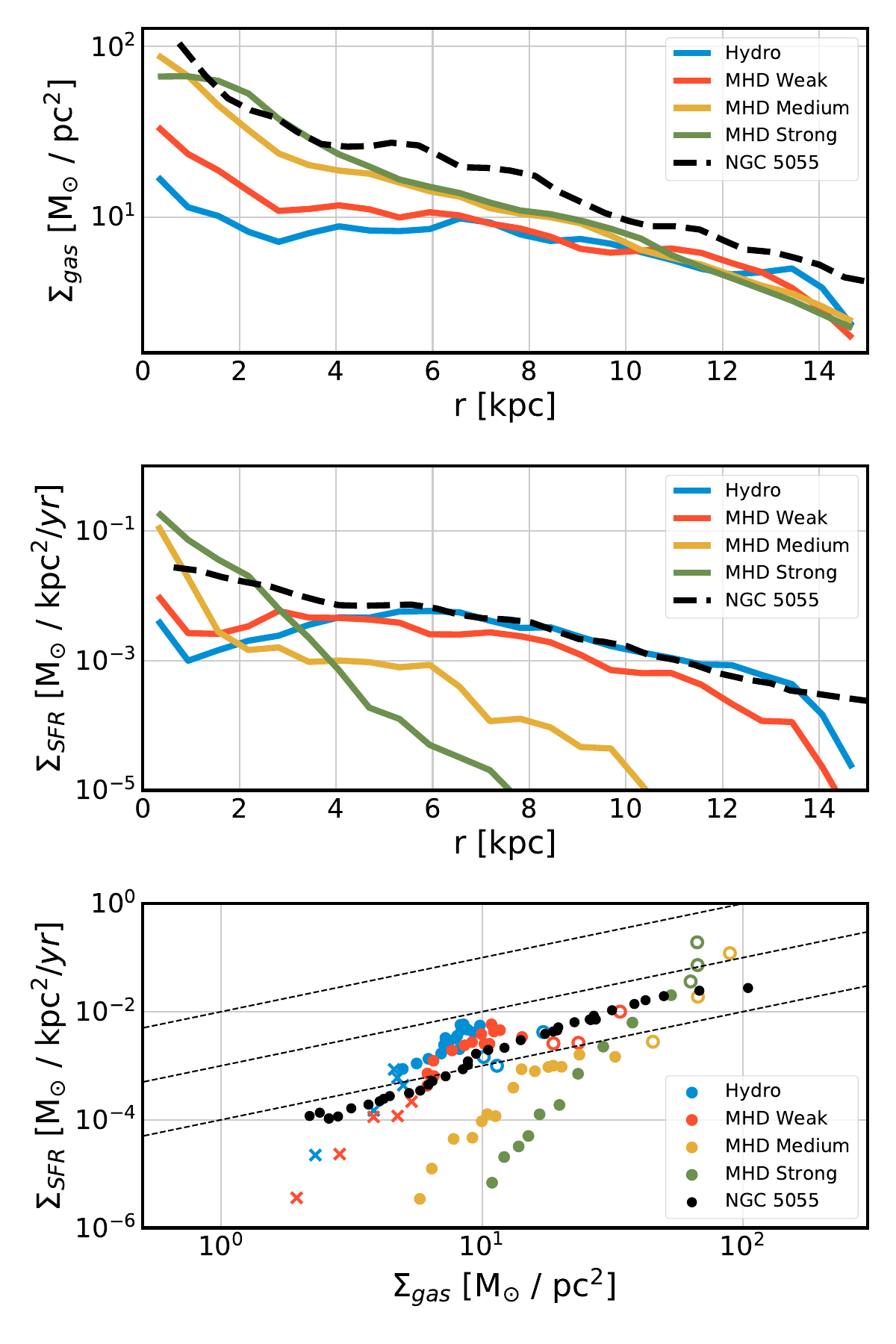}
    \caption{Top row: gas surface density as a function of galactocentric radius at 1 Gyr. Middle row: star formation rate surface density as a function of radius. Bottom row: star formation rate surface density vs. gas surface density, otherwise known as the Kennicutt-Schmidt relation. Open circles indicate points within 2 kpc of the center, and x's indicate points beyond 12 kpc. Diagonal lines indicate constant gas depletion times of 10$^8$, 10$^9$, and 10$^{10}$ years, as done in \citep{bigiel2008}. Data for NGC 5055 is from \citet{Leroy2008}. }
    \label{ks}
\end{figure}

\subsection{Star Formation}
\label{section:sf}

 Figure \ref{sfr} summarizes the star formation history of each galaxy. To ensure a fair comparison between the galaxies we restrict this analysis to gas outside of the central 2 kpc, which minimizes the differences arising from the large difference in gas depletion in the center of the galaxies in different cases. The Hydro and MHD Weak galaxy undergo starbursts that have SFRs reaching up to 20 M$_{\odot}$ yr$^{-1}$. After the initial burst, the hydro galaxy settles into a constant SFR of $\sim 2 M_{\odot} yr^{-1}$ that reduces slightly by the end due to a decreasing gas content. The MHD Weak galaxy has an an initially elevated SFR of 5 M$_{\odot}$ yr$^{-1}$, which continually decreases and reaches 1 M$_{\odot}$ yr$^{-1}$ by the end of the run, lower than the hydro galaxy. The MHD Medium does initially undergo a slight starburst, but it is delayed until $\sim$75 Myr and much smaller. Once it does begin to form stars it quickly reaches a SFR of 4 M$_{\odot}$ yr$^{-1}$ before decreasing even more quickly than the MHD Weak galaxy, and ends up with a SFR of less than 1  M$_{\odot}$ yr$^{-1}$. The decreasing SFR in both the MHD Weak and Medium galaxies is due to the amplification of the magnetic fields,  As they become stronger they become more dynamically important, they limit star formation more effectively. The MHD Strong galaxy has very limited star formation, remaining around 1 M$_{\odot}$ yr$^{-1}$ or less for its entire evolution. 

The differences in the remaining gas content can be seen in the top row of figure \ref{ks}, which plots the surface density versus galactocentric radius for each galaxy, alongside data from NGC 5055 \citep{Leroy2008}. 
The stronger the fields in the galaxy, the more gas remains due to the different star formation history. The differences are most prominent in the galactic center but exist out to around 10 kpc, beyond which all 4 galaxies have similar gas surface densities. The second row of figure \ref{ks} shows star formation surface density as a function of radius, averaged over the last 100 Myr of the simulation. The stronger field galaxies have enhanced star formation in the center 2 kpc due to having more gas remaining at this point in time, The MHD Strong galaxy is the most centrally concentrated due to the strong magnetic support in the outer disk (See figure~\ref{pressure}). In the outer regions, star formation only occurs out to a limited distance. It is truncated at 14 kpc in the Hydro and MHD Weak cases, 10 kpc in the MHD Medium case, and at 8 kpc in the MHD Strong case. 

The final row in figure~\ref{ks} combines above data to make the well-known Kennicutt-Schmidt plot. The surface densities were measured at 1 kpc resolution, and then binned by radius. Star formation rate surface density is calculated using stars that formed within the last 100 Myr in each pixel to make a fair comparison to observational tracers of star formation rate. The dashed lines on the plot show lines of constant consumption times of 10$^8$, 10$^9$, and 10$^{10}$ years from top to bottom. \citet{bigiel2008} label these using efficiencies per 10$^8$ yr of 100\%, 10\% and 1\%, respectively. Open circles indicate the central 2 kpc which is depleted in the Hydro and MHD Weak galaxies, and x's indicate regions beyond 12 kpc where the disk has not had enough time to evolve significantly. We note the regions beyod 12 kpc have no star formation at all in the MHD Medium and Strong galaxies, hence they do not appear on the plot. 

In the bottom row of figure~\ref{ks}, the Hydro and MHD Weak galaxies have most of their gas at surface densities of 1-10 M$_{\odot}$/pc$^2$.  They approach a consumption time of  $2 \times 10^9$ years at 10 M$_{\odot}$/pc$^2$ with a reduced star formation at lower surface densities. The stronger magnetic fields push the turn down further right, with star formation being drastically reduced at low surface densities.  They extend to higher surface densities towards the center of the galaxies. The SFR surface densities in the Medium and Strong galaxies fall far below the observations, which suggests the initial fields are too strong and result in unrealistically low star formation due to the extra magnetic support. This result is reinforced when we explicitly examine gas support in the next section.

 \subsection{Gas Properties and Distribution}
 \label{gas_properties}
\begin{figure}
    \centering
	\includegraphics[width=\columnwidth]{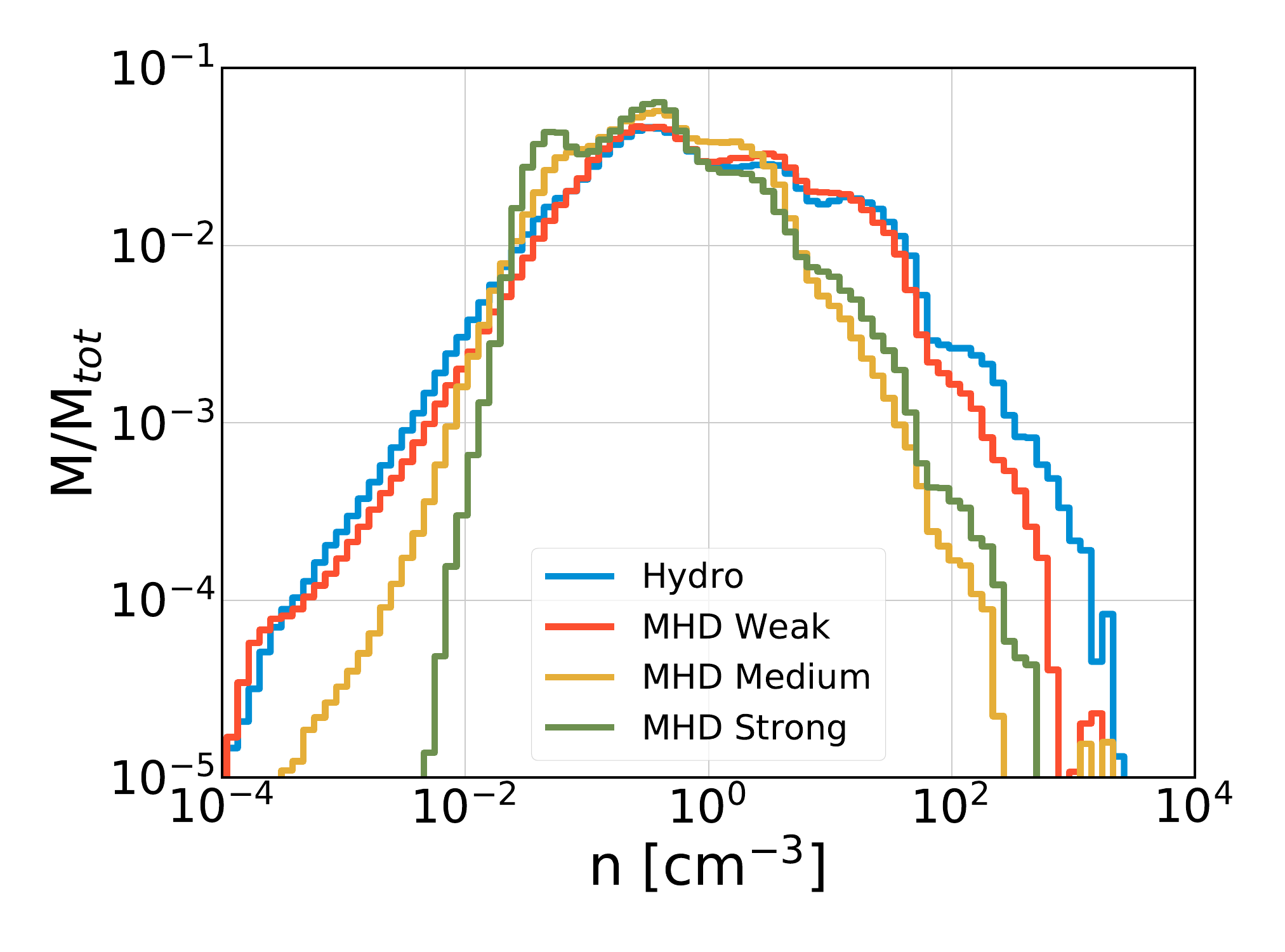}
    \caption{Histogram of gas number densities excluding central 2 kpc. Each bin represents the total mass at that density inside of a disk with radius 15 kpc, and 1 kpc high, normalized by the total gas mass of that disk. Magnetic fields limit the amount of star forming gas that is created by narrowing the distribution.}
    \label{histogram}
\end{figure}

The Hydro and MHD Weak galaxies end up with higher star formation rates than the other two galaxies despite having less gas content remaining at the end.  Our star formation model depends on the amount of dense gas.  Thus we expect to see less dense gas when there is reduced star formation. Figure \ref{histogram} shows a histogram of the total gas mass vs. number density excluding the central 2 kpc at the final snapshot. We see progressively less star forming gas (above the number density of 100) in the stronger field cases as expected. The left side of the distribution is also more extended with the weaker fields. This is because of the stronger feedback and larger bubbles. The regular features separated by factors of 8 in the plot are a result of the refinement strategy of \textsc{Ramses}, with each dip corresponding to a different refinement level.

\begin{figure}
	\includegraphics[width=\columnwidth]{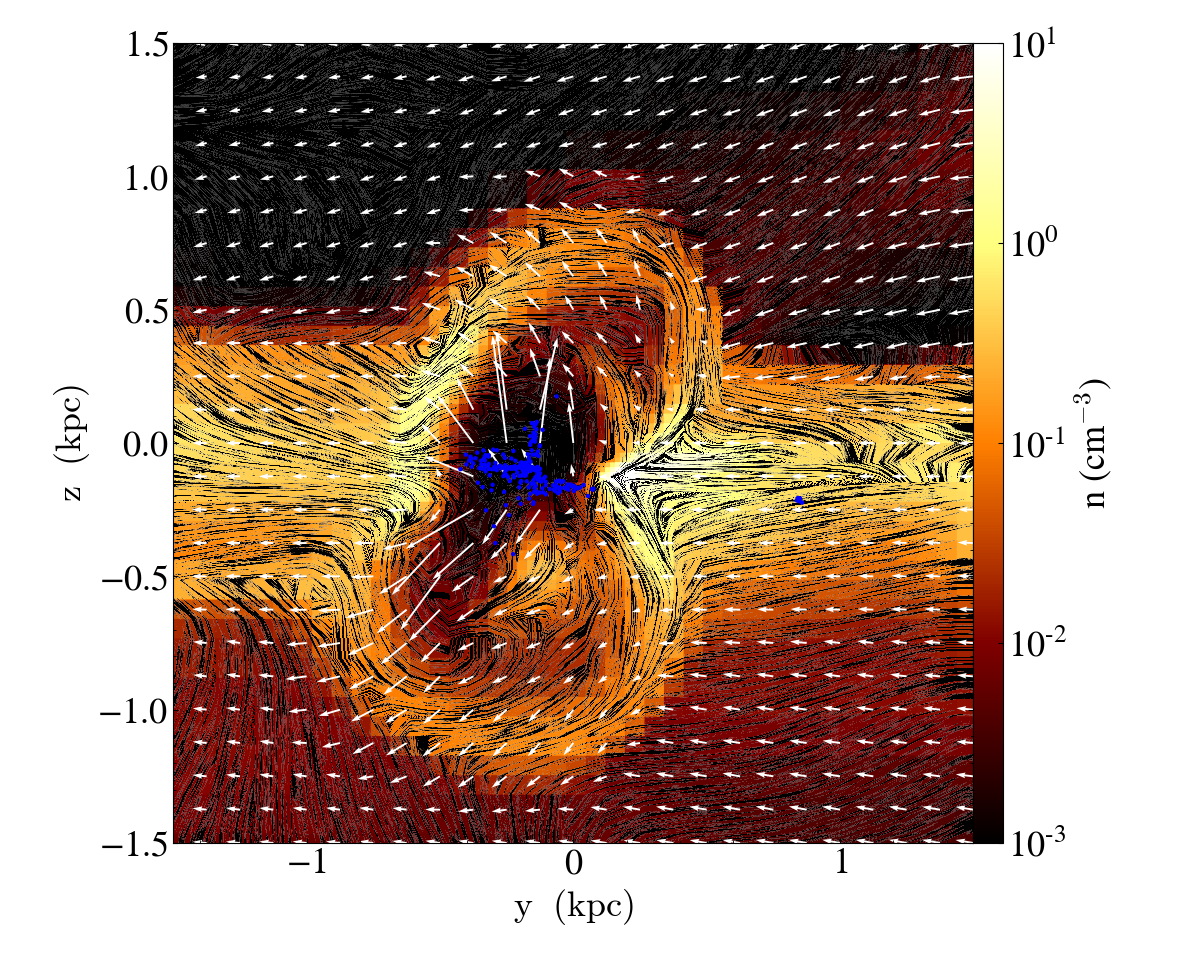}
    \caption{Example of a typical supernova bubble forming in the simulation. Color shows the number density of gas, with the magnetic field lines visualized by a line-integral convolution. Overplotted are the velocity of the gas (white), and newly formed star particles (blue). This particular bubble occurs at a radius of 7 kpc, at t=500 Myr in the MHD Medium galaxy. As the bubble expands, magnetic fields lines are dragged with the gas, and the magnetic tension resists expansion.}
    \label{bubble}
\end{figure}

As individual superbubbles form due to supernova feedback, the magnetic fields lines are dragged by the gas in the explosions, draping themselves around the bubbles. Figure \ref{bubble} shows a typical example of such an event. This particular bubble has stopped expanding and does not end up escaping out of the disk. These explosions are a major source of turbulence in the gas and play a major role in the evolution of both the galaxy and its magnetic fields. 
Magnetic tension causes magnetic fields to resist being bent, counteracting the expansion of the bubbles. Thus the field strength also affects the visual morphology by limiting the number and size of holes in the gas distribution as seen in figure~\ref{projection} and figure \ref{toomre}.
The sizes of the bubbles are also affected by the local surface density. In the MHD Medium and Strong galaxies, star formation is limited to the central region where the high density surrounding medium will also limit the size of the bubbles. However the MHD Weak and Hydro galaxy show show different size bubbles despite having SFRs that are similarly distributed, demonstrating how magnetic fields limit bubble sizes.

\begin{figure*}
    \centering
	\includegraphics[width=2\columnwidth]{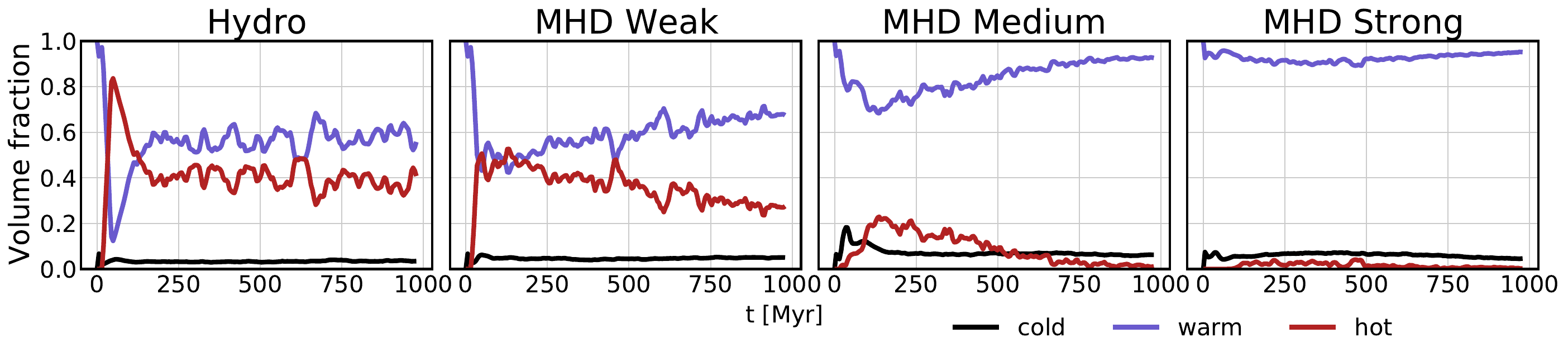}
    \caption{Volume fraction of each phase of gas in the ISM over time. Volume is calculated in a disk within a radius of 2-15 kpc, and height of 1 kpc}. Majority of the volume is split between the warm and hot phase gas. With increasing field strength the volume fraction of the warm phase gas increases, and hot phase gas decreases. 
    \label{volume}
\end{figure*}

The volume fractions of each temperature phase vary for each case, as shown in figure~\ref{volume}. We define the three phases as cold gas (T < 5000 K), warm gas (5000 K < T < 50000 K), and hot gas (T > 50000 K) and exclude the central 2 kpc, just as for the SFR plots. To first order, the volume fractions are explained by higher star formation rates leading to more supernovae and more hot gas in superbubbles. As the star formation rates decrease in the MHD Weak and Medium galaxies, the volume fraction of hot gas decreases correspondingly.  In addition, cases with stronger magnetic fields have a lower hot volume fraction for a given star formation rate. Between 500 -700 Myr, the MHD Weak galaxy has roughly the same star formation rate as the Hydro galaxy, but a systematically lower hot volume fraction. Similarly, when the MHD Medium galaxy peaks in star formation around t=150 Myr, which is higher than the hydro galaxy, the hot volume fraction peaks at just over 20 percent. The MHD Strong galaxy has few, small bubbles. Its volume is almost entirely dominated by warm gas. The MHD Medium galaxy ends up in a similar state by the end of the simulation. 

This strongly suggests that the magnetic fields are limiting the growth of superbubbles, as seen in figure \ref{bubble}. Another possible cause of the difference in bubble volume is the clustering of stars, if the star formation is more clustered, the supernovae combine more efficiently and will grow even larger \citep{nath2013,keller2014}. We have confirmed that the masses of the star clusters do not change between the four galaxies, all having similar distributions.
It is commonly estimated that hot gas occupies $\sim 50$ \% of the ISM by volume \citep{tielens2010}.  Only the MHD Weak and Hydro cases reflect this.  The MHD Medium and Strong filling factors in figure~\ref{volume} extremely low at late times.

\begin{figure}
	\includegraphics[width=\columnwidth]{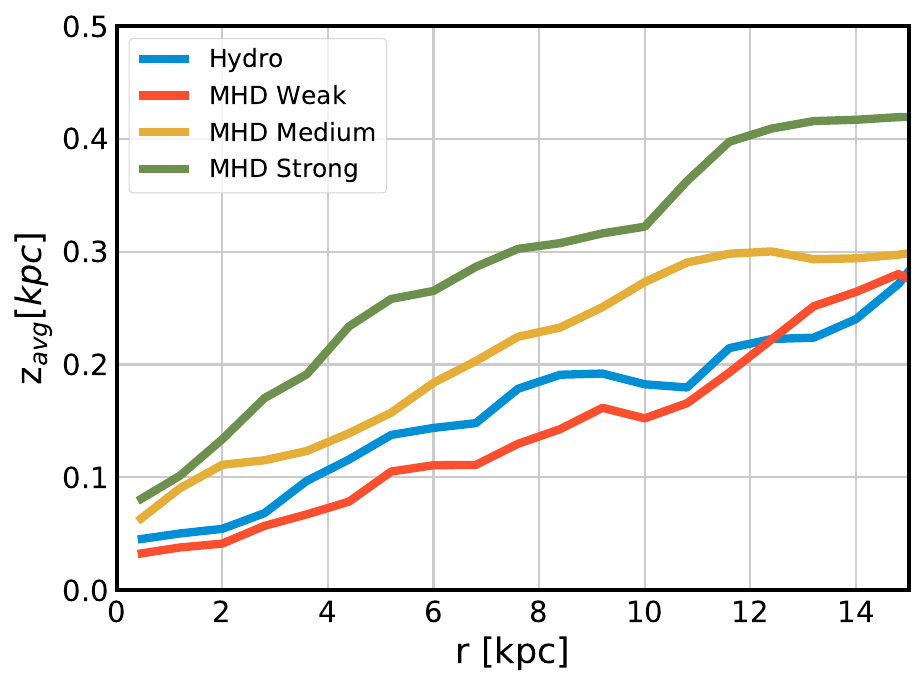}
    \caption{Mass weighted average |z| (which represents the thickness of the disk), as a function of galactocentric radius. In each radial bin, gas within 4 kpc vertically of the midplane was included. The MHD Weak galaxy has a thinner disk than the hydro case, but the more strongly magnetized galaxies are thicker.}
    \label{scaleheight}
\end{figure}

Figure \ref{scaleheight} shows the mass weighted average of the height of the gas vs. radius. All of the galaxies have a value less than 100 pc in the center, increasing as the disk flares outwards. The thinnest disks occur in the MHD Weak and Hydro galaxies which both reach a height of 250 pc by a radius of 15 kpc. Of the two, the MHD Weak's disk is slightly thinner, because the small increase in the thickness from the magnetic support was countered by the reduced turbulence as a result of the lower SFR. The MHD Medium and Strong galaxies are both systematically thicker despite their drastically reduced star formation rates. The MHD Strong galaxy is the thickest, reaching a height of 400 pc. Because of the mass weighting this measure of thickness is mostly set by the high density gas, but in the galaxies with stronger fields there is less cold gas so the increased height is largely due to warm diffuse gas which is magnetically supported. 
To quantify these results, we next examine the different contributions to the supporting pressure.

The differences in gas properties and star formation rate ultimately arise due to the different forces (or pressures) acting on the gas, including thermal pressure, magnetic forces, turbulent motions and gravity \citep{benincasa2016}. Galaxies that have stronger magnetic fields have reduced turbulence due to the reduced stellar feedback.  

\begin{figure*}
    % notice difference in turbulent
	\includegraphics[width=2.1\columnwidth]{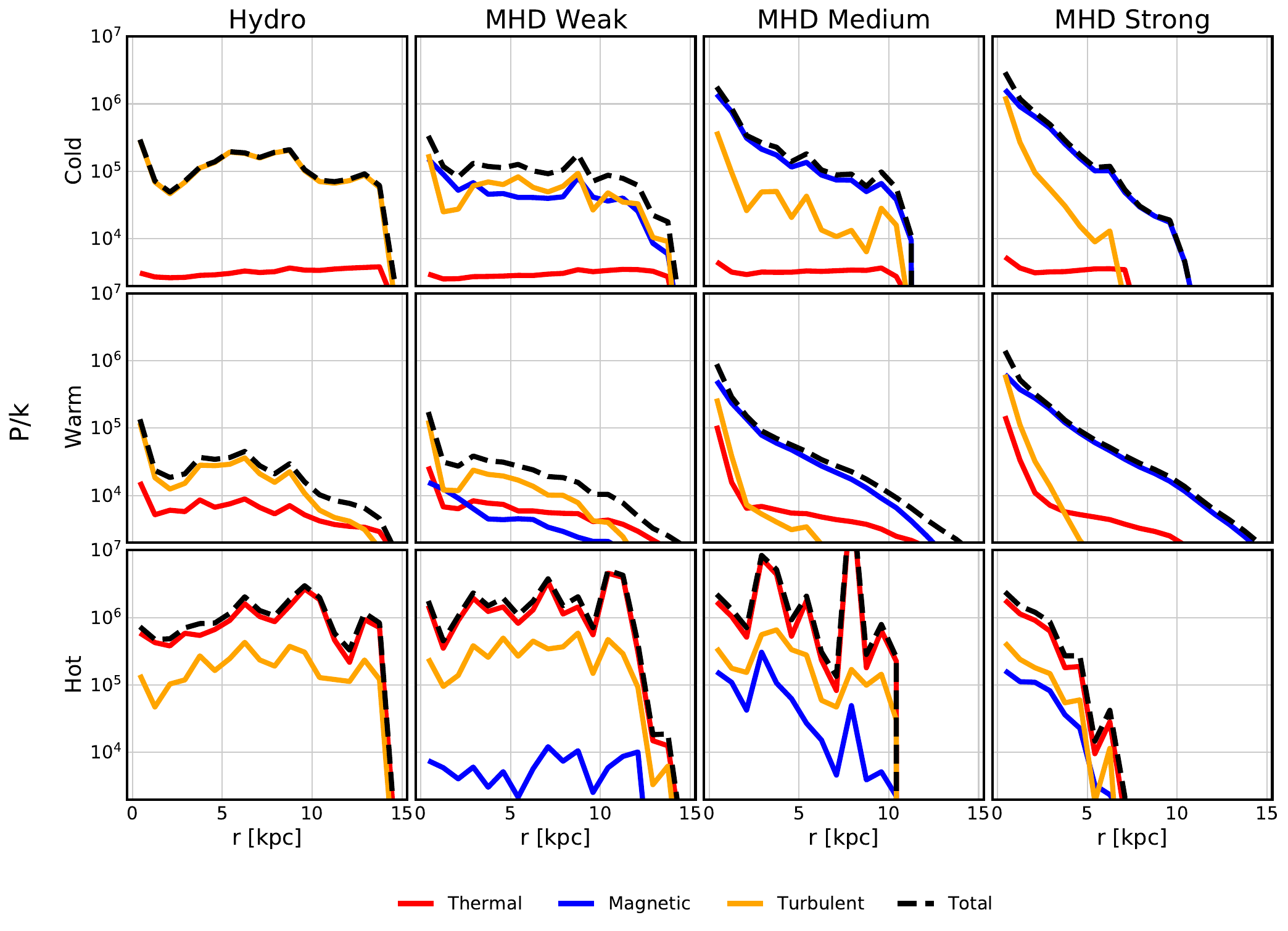}
    \caption{Comparison of different pressures in each galaxy. Included pressures are thermal (red), magnetic B$^2$/8$\pi$ (blue), and turbulent $\rho v_z^2$ (yellow). Each pressure is calculated by taking a mass weighted average of gas within 1 kpc vertically in each radial bin. The pressures are calculated for three separate phases, using only the gas that falls within each temperature range: cold (T < 5000 K), warm (5000 K < T < 50000 K), and hot (T > 50000 K).}
    \label{pressure}
\end{figure*}

Figure \ref{pressure} shows mass-weighted pressures providing vertical support within 500 pc of the disk midplane versus radius, averaged over the final 100 Myr.  The curves shows thermal pressure, $P_\textrm{thermal} = n\,k\,T$, magnetic pressure $P_\textrm{B} = B^2/8\pi$, and a turbulent pressure, estimated using $P_\textrm{turb} = \rho v_z^2$, where $v_z$ is the z-component of the velocity. The top row is cold gas, the middle row is warm gas, and the bottom row is hot gas, using the same definitions as above (i.e. in figure~\ref{volume}). In the cold gas, the MHD Weak, Medium, and Hydro galaxies have roughly the same total pressure, however with increasing magnetic field strength there is a smaller contribution from turbulence. The central regions of the MHD Medium and Strong galaxies have higher support to hold up the extra gas remaining there, and the radial extent of the cold phase is reduced.
We draw attention to the difference between the Hydro and MHD Weak cases.  The turbulent support is roughly half for MHD Weak case (with the difference being made up for by magnetic support), mirroring the halving of the star formation rate at this time relative to Hydro.

In the warm gas, the difference in magnetic pressure is much larger between the MHD Weak and Medium. The increase in magnetic pressure in warm gas is largely responsible for the increased scale height in the Medium and Strong Galaxies. All pressures are in a rough equipartition in the MHD Weak galaxy, but the warm phase becomes dominated by magnetic pressure in the Medium and Strong galaxies.

The hot gas is localized in high pressure superbubbles which are dominated by thermal pressure in all galaxies.  The magnetic fields expand with the hot gas, resulting in the lower magnetic pressure in the hot gas. There is also a high turbulent pressure, which is likely due to high-velocity flows as opposed to small scale turbulence.   

%punch line after this is that we are exploring steady state -- we have shown y axis as energy which is also pressure -- one plot that has magnetic pressure in all 4 cases

\subsection{Disk stability}
\label{stability}
\begin{figure*}
	\includegraphics[width=2\columnwidth]{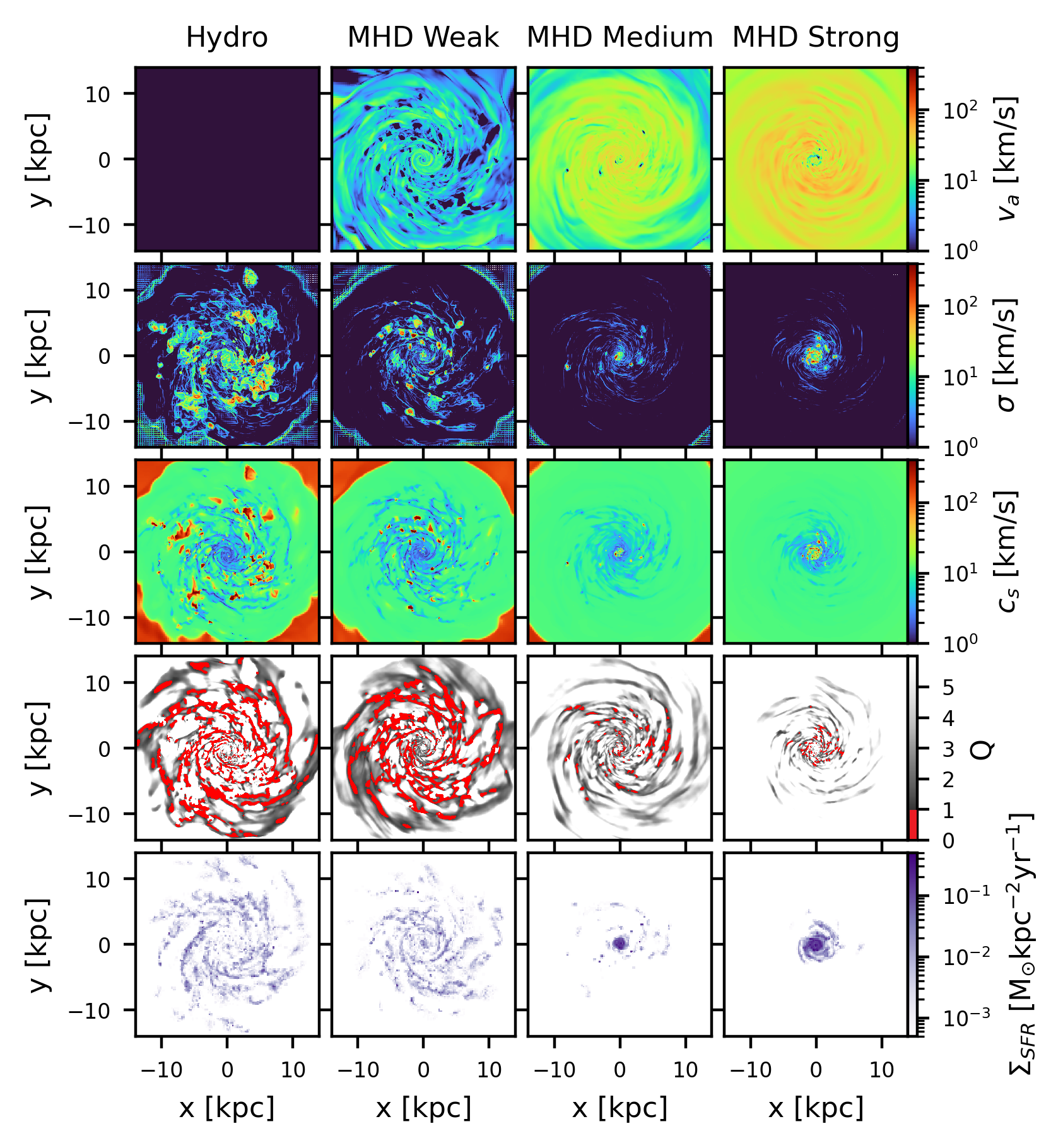}
    % can we make a similar plot to this with mass per unit length / hill radius. To determine difference in fragmentation. Hacar et al 2022, section 2
    \caption{Toomre Q parameter and the three support terms included in it as shown in Equation \ref{toomre_eq}. Top row: Alfvén velocity. Second row, velocity dispersion. Third row: sound speed. Fourth row: Toomre Q parameter. Red regions show Q<1, which are unstable to collapse. Q values of 1-2 are unstable to forming spirals. Fifth row: Surface density of star formation using stars formed within the last 100 Myr.
     }
    %note that these plots look different at different times, especially the MHD medium which has much more red earlier on in its evolution. Is it just the endstate we care about?
    
    \label{toomre}
\end{figure*}

We have shown that the formation of a cold phase and subsequent star formation is much more limited in the stronger field cases.   Gravitational instability drives the formation of the cold phase.  We can quantify this using the Toomre Q parameter for the gas,
\begin{equation}
    Q = \frac{\kappa \sqrt{c_s^2 + v_a^2+\sigma_v^2}}{\pi G \Sigma},
    \label{toomre_eq}
\end{equation}
which accounts for the thermal, magnetic and turbulent support discussed in the previous section. Here $c_s$ is sound speed, $v_a$ is the Alfvén velocity, $\sigma_v$ is the velocity dispersion of the gas, and $\Sigma$ is the surface density of the gas. The literature contains several extended versions of the Toomre Q parameter \citep{romeo2013,Koertgen2019,q_lyon}, which include adjustments for the 3D structure of the disk and multiple components like the stellar disk. We note that the \textsc{Agora} stellar disk has quite a high velocity dispersion and thus the stellar component is quite stable and does not contribute much to the effective Q locally.   The regions of low gas Q closely match up with the locations where stars form, confirming that this choice of Q is a reasonable approximation.   Figure \ref{toomre} shows the Toomre Q parameter for each galaxy at 1 Gyr, along with each of the individual support terms. The Alfvén velocity and sound speed were calculated by taking a mass weighted average of gas within 250 pc of the midplane, and the velocity dispersion in the midplane is estimated by summing over the differences between neighbouring cells. 

The hydro galaxy again has highest turbulent support which is localized around star forming regions.  Conversely, velocity dispersion provides the least support in the magnetic galaxies. Those same regions also contain high temperature gas which yields high sound speeds. The prevalence of hot bubbles drops rapidly as the field strength increase from left to right in the third row of the figure.  The thermal support is lowest in dense gas seen along spiral arms, where the magnetic support is highest. 

The fourth row of figure \ref{toomre} shows the Toomre Q parameter, with red regions indicating Q$<1$. Much of the material with lower Q values has already collapsed so these values below 1 are not informative.  Each galaxy is unstable only in regions of high surface density along spiral arms. The galaxies with stronger magnetic fields have dramatically fewer regions that are unstable and at large radii become totally stable against collapse, explaining the locations of star formation in the fifth row.  Our MHD Weak galaxy has similarly placed but smoother unstable regions when compared to the hydro galaxy.  This is directly related to the addition of magnetic support shown in the top panel. 
The spiral features are thus smoother and more continuous, which is a general feature in the magnetized cases.

As discussed in \citet{Koertgen2019}, magnetized galaxies may also be able to fragment via the Parker instability rather than gravitational instability, potentially allowing  collapse in regions where $Q>1$. This would be difficult to observe in our galaxies because stellar feedback is constantly stirring the gas and may disrupt the wavelike structure required. However, we see relatively little star formation outside of Toomre unstable regions.  In addition, the timescale for these large-scale Parker instabilities tends to be long compared to turbulent crossing times. 

In the outer regions of the disks, much of the gas is marginally stable ($1 \leq Q \leq 2$).  In this regime, the development of spirals is still possible due to swing amplification \citep{binney_tremaine}.  These spirals form but do not necessarily fragment into clouds. In the MHD strong case no dense structure forms in the outer regions (see figure \ref{projection}) due to the enhanced magnetic support, and the star formation is heavily centrally concentrated, as shown in the bottom row.

\section{Discussion}
\label{discussion}
We simulated four galaxies with varying magnetic field strengths in order to understand the impact on their ISM and star formation regulation. 
We saw clear differences between the galaxies in their star formation rates.  This result arises due to the consumption of gas fuel and the growth of the magnetic fields strength over time.  Star formation rates can be understood via the gas support in each case. This is clearly reflected in the stability of the gas.  Stronger fields are able to support the gas without feedback, particularly at large radii.  Weak or absent fields must rely on turbulent support associated with star formation.  Every case is internally self-consistent.  However, the combination of the star formation rates, field strengths and gas morphology heavily favour the steady state generated by the MHD Weak case as the best match to observed disk galaxies.

A lack of star formation also means a lack of cold, dense gas precursors for star formation, which is reflected in the ISM gas phase results.  This is tied to the Schmidt-law star formation model, that always forms stars at a fixed efficiency per free-fall time once gas is above a density threshold.  
While this is a standard model choice, it is important to recognize that the actual efficiency of star formation in dense gas is the subject of vigorous theoretical and observational study.  In particular, the presence of unresolved magnetic fields and turbulence could affect this efficiency in real galaxies, whereas the efficiency is fixed here \citep{Federrath2015}.

It is difficult to compare simulated field strengths to observed values.  We argue that volume weighted fields are similar to observed synchrotron estimates.  In particular,  the MHD Weak case creates a flat radial profile.  While this is compelling relative to the other cases,  mock observations would provide more detailed insights \citep{ponnada_2022,ponnada2023,martin2024}. 

The MHD Weak galaxy produced a steep power law increase in field strength at high densities.   This is reminiscent of the upper envelope of the Zeeman observations \citep{crutcher}. However, our ideal-MHD simulated field strengths should be upper limits, as turbulent ambipolar diffusion will reduce the field at high densities \citep{heitsch2004}. Our simulations  also show no evidence for the commonly inferred constant field strength in lower density gas. The lack of constant field has been observed by several independent simulation groups \citep{Koertgen2019,Rieder3,ponnada_2022}.  Thus there is a tension between observations and theory on this point.

It is tempting to ask which of the four galaxies are the most realistic.  Magnetic fields certainly exist in galaxies, ruling out the hydro galaxy.
Similarly, the MHD Strong galaxy had initial fields above what we infer from observational constraints.  Both it and the MHD medium case failed to match most observed properties for the field itself, gas morphologies and star formation. 

The MHD Weak case did well on all these measures.
We expect that any case starting with a sufficiently high plasma beta ($\gtrsim 100$) would evolve to a saturated
state similar to the MHD Weak case.  The main difference would be the time it takes to do so.  The key requirement is
that the fields are allowed to amplify naturally to the point of saturation. 

We note that the MHD Weak galaxy achieves similar star formation rates and gas distributions to the Hydro galaxy at intermediate times.  Eventually, the gas support includes a magnetic component similar to the turbulence and
the star formation rate dips below the hydro case.  
In future work, we will explore the morphology of the field and how it is linked to its ability to support the gas in place of stellar feedback. This may be related to the split between turbulent and mean fields. It would also be worth testing different morphologies in the initial condition. We started with a purely toroidal field which could possibly bias the final field configuration.

%add section comparing to other work
This work adds to the growing list of MHD galaxy simulations. \citep{wang2009, Pakmor2013,martin2020,wibking2023}. Similar to other simulations of isolated disks, there is initially fast magnetic field amplification due to an central starburst, followed by a self-regulated amplification phase as the fields approach saturation. The self-regulated phase is consistent with the feedback-driven amplification seen in late times in cosmological simulations, and produces similar field strengths \citep{martin-alvarez2018,pakmor}. Several isolated galaxy studies have also demonstrated that supernovae play a major role in field amplification, which our results are consistent with \citep{butsky2017,Rieder1}.  It has also been found that magnetic fields reach equipartition with thermal energy once saturated \citep{pakmor2014}. Our results show that happens specifically in warm-phase gas, while in cold gas the fields are above equipartition, and in hot gas they are below equipartition. Our findings that most of the amplification occurring in the warm phase has also been seen in simulations by \cite{martin2022}.

This work does not account for cosmic rays, which could contribute significantly to gas support in principle (e.g. \citealp{semenov2021}).  We note that there is a lot of uncertainty regarding cosmic rays coupling to the gas and how best to model them numerically.  While the current work uses simple approaches to isolate key effects, cosmic rays would be an interesting future direction.

We also do not account for non-ideal MHD effects. While ideal MHD is a good approximation for most of the ISM, effects such as ambipolar diffusion may become important in high density star forming regions.  We anticipate that these effects are comparable to the impact of the magnetic field on small star formation generally, which is not addressed in our simple star formation prescription.   We also avoided more elaborate feedback models.  Complex feedback models would also be affected by unresolved turbulence and magnetic fields.  Though we chose simple feedback by design to keep interpretation simple, it is worth keeping in mind that factors like star formation efficiency could be different in different cases (e.g. with stronger magnetic fields).

Studying the properties of star forming regions and molecular cloud analogues is a natural extension of this work to smaller scales.  Magnetic fields may affect the shapes and sizes of star forming clouds, which could also affect small-scale star formation.  To do this, we need higher resolution simulations. We have performed zoom-in simulations, beginning with the  galactic setup in this work, seeking to characterize the affect of galactic-scale dynamics on star forming regions (Zhao et al 2023, in prep).

Another population of interest within galaxies is that of superbubbles. JWST observations have made it possible to perform a detailed census of superbubbles in nearby galaxies \citep{watkins2023}. Our results show that magnetic fields dramatically affect the evolution of superbubbles.  It would be interesting to characterize their populations in these and higher resolution simulations as a further constraint on how well the simulated magnetic fields match those in real galaxies. 

\section{Conclusions}
\label{conclusions}

We have performed simulations of magnetized Milky-Way like galaxies with different magnetic field strengths. As the galaxies evolved differences in gas distributions and star formation occur due to the differing levels of magnetic support which in turn result in different levels of turbulent support from the stellar feedback. We summarize our conclusions as follows:

\begin{itemize}
  \item We have demonstrated that evolving a simulated galaxy towards a realistic self-regulating ISM with magnetic fields is achievable with an initially weak toroidal field and simple models for star formation and feedback. Conversely, starting with field strengths typical for the ISM limits star formation though magnetic support.
  \item Stronger magnetic fields generally reduce a galaxy's star formation. As magnetic fields amplify over time and their strengths increase, star formation rates will decrease.  Field strengths evolve due to the star formation and feedback and provide an additional form of self-regulation in galaxies.
  \item In our isolated galaxies, dynamo amplification mainly occurs in the warm diffuse medium of the ISM, from number densities of 0.1 to 10 cm$^{-3}$. The galactic magnetic fields first saturate at high densities but continue to grow at intermediate densities over longer timescales.  
  
  \item Starting with high initial fields may lock the galaxy into field configurations set by the initial conditions that result in unrealistically low star formation. Stronger magnetic fields generally result in reduced turbulence. This is due to both the reduced supernovae feedback, but also from magnetic fields limiting superbubble growth.  Smaller feedback bubbles can be seen in the projected gas density and by the reduced volume fraction of hot gas for a given star formation rate. 
     \item Magnetic fields change the distribution of unstable regions in the disk, which are well characterized with a modified gas Toomre~Q. Strongly magnetized galaxies become completely stable in the outer disk and star formation becomes more centrally concentrated.
  \item Initially weak magnetic fields evolve to play an important role in the vertical pressure support in galaxies, approaching equipartition levels with turbulence in the cold phase and with both thermal pressure and turbulence in the warm phase. This adds an extra component the established pressure regulation picture \citep{ostriker2010,benincasa2016}.
  \item Initially weak fields  saturate at $\sim 1$ $\mu G$ (volume weighted) in the warm medium at a range of radii, while still producing a steady-state, self-regulated galaxy.  
  While the flat B profile is consistent with synchrotron observations, the $\sim 1$ $\mu G$ amplitude is low.  The amplitudes inferred from observations rely on several assumptions.  Thus it may also be worth exploring if the observed amplitudes are overestimates. 
  \item  The evolved, self-regulated state of the MHD Weak simulation best matches the expected star formation radial profile, taken from either the Kennicutt-Schmidt relation or the nearby galaxy NGC 5055, which the \textsc{Agora} initial condition is similar to.  This provides further support for the idea that lower ($\sim 1$~$\mu G$) magnetic fields at typical ISM densities (0.1-1~cm$^{-3}$) are consistent with realistic galaxies.

\end{itemize}

\section*{Acknowledgements}
The authors would like to thank Ralph Pudritz for many useful dicussions. HR is supported by an NSERC postgraduate scholarship, and 
JW is supported by Discovery Grants from NSERC of Canada.  Computational resources for this project were enabled by a grant to JW from Compute Canada/Digital Alliance Canada and carried out on the Niagara Supercomputer.
%%%%%%%%%%%%%%%%%%%%%%%%%%%%%%%%%%%%%%%%%%%%%%%%%%
\section*{Data Availability}

The data used this article will be shared upon reasonable request to the corresponding author.



\bibliographystyle{mnras}
\bibliography{example} % if your bibtex file is called example.bib

%\begin{thebibliography}{99}
%\end{thebibliography}

%%%%%%%%%%%%%%%%%%%%%%%%%%%%%%%%%%%%%%%%%%%%%%%%%%

%%%%%%%%%%%%%%%%% APPENDICES %%%%%%%%%%%%%%%%%%%%%

%\appendix



%%%%%%%%%%%%%%%%%%%%%%%%%%%%%%%%%%%%%%%%%%%%%%%%%%

% Don't change these lines
\bsp	% typesetting comment
\label{lastpage}
\end{document}